\documentstyle[14pt]{article}
     \newtheorem{pp}{Corollary}
\begin{document}
\begin{center}
     {\large\bf About the Poisson Structure\\
for D4 Spinning String
\footnote{to be published in Journal of Physics A}}\\[5mm]
     S.V. Talalov\\
     {\small Dept. of Theoretical Physics,  University
      of Togliatti, \\
    13 blvrd Korolyova, Togliatti, 445859, Russia}\\
e-mail svt@tfsgpu.edu.tlt.ru \\[5mm]

\begin{abstract}
     The model  of D4 open  string with non-Grassmann spinning
     variables is considered. The non-linear gauge, which is
     invariant both Poincar\'e and scale transformations of the
     space-time, is used for subsequent studies.
     It is shown that the reduction of
     the canonical Poisson structure from the original phase space
     to the surface of constraints and gauge conditions
     gives  the degenerated Poisson brackets. Moreover it is
     shown that such reduction is non-unique.
     The conseption of the adjunct phase space is introduced.
     The consequences for subsequent quantization are discussed.
     Deduced dependence of spin $J$
     from the square of mass $\mu^2$ of the string  generalizes the
     ''Regge spectrum`` for conventional theory.

     \end{abstract}
     \end{center}

     \subsection*{1. Introduction}

     The investigation of the constrained dynamical systems
     was started by Dirac
     \cite {Dir}
     and is continued in connection with the gauge theories
     development. There are many directions of the studies
     exist here; one of them is  the
     modification of the conventional phase space conseption
     (see, for example, \cite {Lus}).
     In this article, firstly, we suggest new viewpoint
     on the phase space for some kind of the gauge systems and,
     secondly, we apply the suggested consepts to investigation
     of D4 string dynamics.

     We start our studies with the following simple example.
     Let the  space  ${\cal H}_N$ be the phase space of
     any dynamical system with $N$ degrees of the freedom;
     any point $M\in{\cal H}_N$ has the coordinates
      $p_1, q_1,\dots,p_N, q_N$ which diagonalize the
      standard non-degenerated Poisson brackets: $\{p_i,q_j\}=
     \delta_{ij}$, $\{p_i,p_j\}=\{q_i,q_j\}=0.$
     Let us consider the subset $V\subset{\cal H}_N$:
     $V=\{M\in{\cal H}_N~\vert~ p_1=0,~q_1>0\}.$ What is the
     Poisson structure of the set $V$? It is clear that such structure
     must be degenerated because ${\rm codim}V=1.$
     The simplest foliation of the set $V$ will be following:
     $$V=\mathop\cup_{c>0}V^0_c,$$
     where set $V^0_c=\{M\in{\cal H}_N~\vert~
      p_1=0,~q_1=c,~(c=const)\}.$
     It is well-known fact that the ''correct`` brackets for
     any set $V^0_c$ will
     be the Dirac brackets $\{\cdot,\cdot\}_1$ for the pair (second
     type) constraints $p_1=0$, $q_1-c=0.$  This bracket structure
     can be naturally extended on set $V;$ the function
     $f_{0}\equiv q_1$  will be annulator.
     The interesting fact is that
     the constructed  brackets are non-unique. Indeed, we can introduce
     the other foliation of   the   set   $V:$
     $V=\cup_{c>0}V^f_c,$
     where the subsets  $V^f_c=\{M\in{\cal H}_N~\vert~ p_1=0,~
     f(q_1; q_2, p_2,\dots)=c,~(c=const)\}$ were defined with help of
      some appropriate function $f$ such that condition
     $0<{{\partial f}\over{\partial q_1}}<\infty$ holds.
     It is clear that corresponding  Dirac brackets $\{\cdot,\cdot\}_f$
     differ from the brackets  $\{\cdot,\cdot\}_1;$ the annulator
     for new brackets is the function $f.$
     Thus, the reduction of same Poisson
     structure from the original phase space ${\cal H}_N$
     on some subset $V\subset{\cal H}_N$ can be ambiguous if the
     reduced brackets degenerated.
      In general, the situation is same if we
     consider the system of the first type constraints
     $f_1,\dots,f_k,$ where $k<N,$ instead the single one
     $p_1=0.$    Of course, this example is the
     special case of the general theory of degenerated Poisson
     brackets
     \cite{KarMas}.
     It was discussed in detail because the goal of our subsequent
     studies will be investigate this effect in a string theory.
                     $$***$$

     The  satisfactory version of D4 quantum
     (super)string was a purpose  of the theoretical studies for many authors
     (see, for example, \cite {Roh, Mar2, Chams, Pron, Lun}).
     Moreover, many authors construct the theories in arbitrary
     (non-critical)  space-time dimensions
     \cite {Pol, GerNev, Mar1}. Of couse, this list is uncomplete: the
     detail review is impossible here.
     Suggested  approach differs, in our knowledges, from other
     because it founded on the new conception of adjunct phase space.

     So, we consider the following  model here. Let
     the fields $X_\mu(\xi^0,\xi^1)$ and
     $\Psi^A_{\pm}(\xi^0,\xi^1)$
     interact with two-dimensional gravity
     $h_{ij}(\xi^0,\xi^1),$
    where $\xi^1\in[0,\pi]$ and $\xi^0\in(-\infty,\infty),$
    such that dynamics is defined by the action
      constructed in accordance with the
     well-investigated manner
     \cite{GSW}:

     \begin{equation}
     S=-{1\over{4\pi\alpha^{\prime}}}\int d\xi^0d\xi^1\sqrt{- h}
     \left\{h^{ij}\partial_i X^\mu\partial_j X_\mu -
     i\Theta e^j_I(\Gamma^0)_{AB}\overline\Psi^A\gamma_j
     {\nabla^I}\Psi^B\right\}.
     \end{equation}

     The notations are  following:    $h=\det(h^{ij})$,
     the vectors  $e^j_I(\xi^0,\xi^1)$  are the vectors of two-dimensional
     basis such that the equalities  $h^{ij} = e^{iI}e^j_I$ take place
     and the matrices  $\Gamma^\mu$  and $\gamma_i$ are the Dirac
     matrices in the four- and two-dimensional space-time respectively.
     The field  ${\bf X}=X_\mu{\bf t}^\mu$
     is the  vector field  in (''isotopical``)
     Minkowski space-time  $E_{1,3};$  the fields  $\Psi^A$
     with components $\Psi^A_\pm$ are the spinor fields in two-dimensional
     space; index $A$ is the spinor index in the space $E_{1,3}$
      such that the fields  $\Psi_\pm$ are the Majorana spinor
     fields in four-dimensional space-time.
     The numbers $\Psi^A_\pm$ are the complex numbers, so
     there are no classical Grassmann variables in our action.
     The consideration of the spinning string without the
     Grassmann variables does not new (see, for example,
     \cite {Bar}).
     In our opinion such approach is justified here because
     the  new fundamental variables  will be the
      complicated functions from the original fields $X$ and
     $\Psi.$

     To fix the gauge arbitrariness  we demand, as usually,
     $e^j_I=\delta^j_I$ so that $h_{ij}={\rm diag}(1,-1)$ and
     the equations of motion can be written in simplest form.
     For fields $X$ and $\Psi$ we have
$\partial_-\partial_+X_{\mu}=0,$
$\quad\partial_{\mp}\Psi_{\pm}=0;$
     the equations of motion $\delta S/\delta h^{ij}=0$
     for gravity $h$ lead as well-known to the equalities
     \begin{equation} F_{1\pm}(\xi)\equiv
     \left(\partial_\pm X\right)^2\pm
     {{i\Theta}\over 2}\overline\Psi_\pm\partial_\pm\Psi_\pm=0,
     \end{equation}
     where $\partial_\pm$ are derivatives with respect to
     cone parameters $\xi_\pm=\xi^1\pm\xi^0.$
      The remained gauge freedom
\cite{GSW}
     \begin{equation}
     \xi_\pm\longrightarrow\tilde\xi_\pm=\pm A(\pm\xi_\pm),
     \end{equation}
     must be fixed by means of additional conditions.
     For our subsequent consideration it is important
    that the  function $A(\xi)$ must satisfy  the property
    $$A(\xi+2\pi)=A(\xi)+2\pi,\qquad A^\prime\not=0$$
     in accordance with the standard boundary conditions for
     original variables $X$ and $\Psi$:
$ X^{\prime}_{\mu}(\xi^0,0)= X^{\prime}_{\mu}(\xi^0,\pi)=0,$~
$\Psi_+(\xi^0,0)=\Psi_-(\xi^0,0)$ and
$\Psi_+(\xi^0,\pi)= \epsilon\Psi_-(\xi^0,\pi),$
      where $\epsilon=\pm.$

      The original phase space ${\cal H}$ has the coordinates
     $\dot{X}_\mu\equiv\partial_0 X_\mu$, $X_\mu$,
      ${\mathop{\Psi^+}^A}_\pm$ and $\Psi^A_\pm.$ As usually,
     canonical Poisson bracket structure is following:
     $$\{\dot{X}_\mu (\xi),X_\nu (\eta)\}=-4\pi\alpha^\prime
     g_{\mu\nu}\delta(\xi-\eta),$$
     $$\{{\mathop{\Psi^A}^+}_\pm(\xi),\Psi^B_\pm(\eta)\}=
     {8\pi i\alpha^\prime\over{\Theta}}(\Gamma^0)^{AB}\delta(\xi-\eta).$$

     \subsection*{2. The additional gauge conditions\\
      and the   adjunct phase space.}

     The spinor  variables give the additional possibilities
     to construct the natural Poincar\'e-invariant structures on the
     $(\xi^0,\xi^1)$-plane
     \footnote{not only first and second quadratic form of the
     world-sheet, as in the case of bosonic string}.
     For example, we can construct
     the following two-tensor:
     $$\Omega_{ij}(\xi^0,\xi^1)=
     {1\over 2}\left(h_{im}h_{jn}+h_{in}h_{jm}-h_{ij}h_{mn}\right)
     \left(\Gamma^0\Gamma^\mu\right)_{AB}\overline\Psi^A\gamma^m
     \Psi^B\partial^n X_\mu. $$
     Another objects can be constructed too. The detail investigations
     these structures and the geometrical properties of the ''extended``
     world-sheet
     $(\xi^0,\xi^1)\rightarrow
     (X_\mu(\xi^0,\xi^1),\Psi^A_\pm(\xi_\pm))$ in some complex
     space,
     probably, will be interesting, but lie outside the frameworks
     of this article.
     We include  in  our   subsequent  studies   the string
configurations $({\bf X},\Psi)$ which give the positive-defined
     quadratic form $\Omega_{ij}d\xi^i d\xi^j$ only.
     This demand means that two inequalities
     \begin{equation}
     \pm\overline{\Psi}_{\pm}\Gamma^{\mu}\Psi_{\pm}
     \partial_{\pm}X_{\mu}>0
     \end{equation}
     hold for any point $(\xi^0,\xi^1).$
     To destroy the gauge freedom (3) we select
     the string configurations $({\bf X},\Psi)$
     such that the conditions
\begin{equation}
     F_{2\pm}(\xi)\equiv
\overline{\Psi}_{\pm}\Gamma^{\mu}\Psi_{\pm}\partial_{\pm}X_{\mu}
=\pm{\kappa^2\over 2}
\end{equation}
     hold for any non-zero constant $\kappa=\kappa[{ X},\Psi].$
     Note that the equalities (5) are
     invariant both under Poincar\'e and under scale transformations
     of the space-time $E_{1,3},$  so we assume that the resulting theory
     will be  attractive.  Such invariance is first point of the
     motivation for the conditions (5).  Second point is to the  gauge
     (5)  generalizes naturally the well-known light-cone gauge in a
     string theory. We discuss this fact detaily in the end of section
     4.

     It should be stressed that the
     restriction
     \begin{equation}
     \kappa[X,\Psi]=q,
     \end{equation}
     where $q$ is some fixed in-put parameter
     is not suitable for complete theory.
       Indeed, the different values of the constant $\kappa$
     correspond to different orbit of the gauge transformations (3),
     so that $\kappa$ is Teichm\"uller-like parameter. Consequently,
     the ''strong`` restriction (6) will be not grounded because
     the gauge transformations (3)
     were forbidden by the ''weak`` conditions (5)
     (the discussion of this situation for  general gauge systems
     can be found in the work \cite{Fil}).

     Note that the gauge (5) does not forbid the transformations
     (3) such that $A(\xi)\equiv\xi+c,$ where $c=const.$ Obviously,
     they give the shifts $\xi^0\rightarrow\xi^0+c,$ which
     correspond to dynamics.

      We are going to study the Poisson structure of the
     set ${\bf V}$ of string configurations $({\bf X(\xi^0,\xi^1)},
     \Psi(\xi_\pm))$ which are
     selected by the constraints (2) and ''weak`` gauge conditions (5).
     It is non-trivial problem, because the variation
     $\delta(\kappa [X,\Psi])$ does not defined by the variations
     of the coordinates of original phase space.
     Let us introduce the auxiliary minimal subspace ${\cal H}_1$ such
     that, firstly, the inclusion
     ${\bf V}\subset{\cal H}_1\subset{\cal H}$ holds and, secondly,
     the Poisson structure on ${\cal H}_1$ is well-reduced from
     the original phase space
     ${\cal H}.$  Such subspace is given by the equalities
     \begin{equation}
     F^{(n)}_i=0,\qquad n\not=0,\qquad i=1,2.
     \end{equation}
     The constants $F^{(n)}_i$ are Fourier modes of $2\pi$-
     periodical functions
     $$F_i(\xi)=\cases{F_{i+}(\xi),& $\xi\in[0,\pi)$,\cr
     F_{i-}(-\xi),& $\xi\in[-\pi,0)$,\cr}$$
     which are well-defined in accordance with the boundary
     conditions for the variables $X$ and $\Psi.$
     The canonical Poisson structure on original phase space ${\cal H}$
     gives the following brackets:
     $$\{{\mathop{F^{(n)}}^*}_1,F^{(n)}_2\}=8\pi i\alpha^\prime
     n F^{(0)}_2.$$
     Because $F^{(0)}_2=\kappa^2/2>0$
     in our theory, the system (7) will be second type
     system of constraints so that natural brackets on space
     ${\cal H}_1$ will be corresponding Dirac brackets
     $\{\cdot,\cdot\}_1.$  The condition
     $  F^{(0)}_1=0$
     gives the reduction on set ${\bf V};$ obviously,
     ${\rm codim}{\bf V}=1.$   Analogously with
     the example, considered in the introduction,
     we can select the various  foliations
     \begin{equation}
     {\bf V}=\mathop\cup_{q^2>0}{\bf V}^f_q,
     \end{equation}
     where the sets ${\bf V}^f_{q}\subset{\bf V}$ can be defined
     both by the  restriction (6) and any more complicated conditions.
     As for the finite-dimensional case, any foliation (8) gives the
     Poisson structure on the set ${\bf V},$ which will be degenerated.
     Thus, the natural canonical structure
      of the original   phase space ${\cal H}$ does not
     have the unique reduction to the set ${\bf V}.$

     At first it seems that such indeterminacy can be ignored at the
     subsequent quantization. Indeed, we can quantize the brackets
     $\{\cdot,\cdot\}_1$ and construct the correspondent Fock space
     ${\bf H}_1.$ After that we must
     select the physical vectors $\mid\psi\rangle\in{\bf H}_1$
     which will be the solutions of the ''Shr\"odinger
     equation``  $F^{(0)}_1\mid\psi\rangle=0$
     \cite {Dir}.
     In our opinion, the ambiguity in determination of the
     Poisson structure of the manifold ${\bf V},$ which consists
     all physical information, leads to additional
     possibilities for quantization. Indeed, let any space
     ${\cal H}^{ad}$ be any Poisson manifold with the Poisson brackets
     $\{\cdot,\cdot\}^0.$ Suppose that the finite number of some
      constraints    $\Phi_i(\dots)=0$, $i=1,\dots,l$  give
     the first type system of constraints:
     $$\{\Phi_i,\Phi_j\}^0=C_{ijk}\Phi_k.$$
     Suppose, that for the surface of these constraints
     ${\bf W}\subset{\cal H}^{ad}$:
     ${\bf W}=\{M\in{\cal H}^{ad}~\mid~\Phi_i=0,~i=1,\dots,l.\}$
     the diffeomorfism ${\bf V}\approx{\bf W}$
     takes place
     \footnote{this diffeomorfism must be conserved in the
     dynamics}.
     Thus, we have the following diagram:
     \begin{equation}
     {\cal H}_1\supset{\bf V}\approx
     {\bf W}\subset{\cal H}^{ad}.
     \end{equation}
     We call any such space ${\cal H}^{ad}$ the
     {\it adjunct phase space}.
     From the classical viewpoint, the manifold ${\bf W}$
     is tantamount to the manifold ${\bf V},$ because it
     has same information about the physical degrees of the
     freedom.
     As discussed above, the single-defined Poisson structure
     is absent both on the manifold ${\bf V}$ and the manifold
     ${\bf W},$ that is why there are no reasons prefer one to
     other.  Consequently, we can fulfill the subsequent
     quantization of the theory in terms of the space ${\cal H}^{ad}.$
     It should be stressed that ${\cal H}_1\not\approx{\cal H}^{ad}$
     as the Poisson mahifolds in general, even for case $l=1.$
     This means  that the  results
     here can differ from the conventional ones.
     The simplest example of adjunct phase space will be
     the manifold $D*{\cal H}_1,$ where symbol $D*$ means any
     (not only Poisson) diffeomorfism.
     It is clear from the physical point of
     view. Indeed, if we have both physical and  non-physical
     coordinates in some phase space, there are no reasons
     to consider only Poisson-conserved transformations of
     such space. The suggested definition of the space
     ${\cal H}^{ad}$ is more general, of course.

     The main goal of this article is construct the physically
     appropriate adjunct phase space for the dynamical system (1)
     and discuss the quantization.
       In our opinion, there are several reasons to refuse the
     description of string dynamics in term of the
     space ${\cal H}_1$ (or original phase space ${\cal H}$ --
     it is equivalent):
     Firstly,  the standard approach leads to
     the additional  dimensions for space-time while the existence
     of such dimensions is not proven experimentally.
     Secondly, the conventional approaches
     (see, for example, \cite{GSW})
     lead as is well known to the linear Regge trajectories for free
     strings such that the slope $\alpha^\prime$ is in-put parameter
     in a theory.  But
     the   trajectories $s=\alpha^\prime\mu^2+c,$
     where the value $\alpha^\prime\simeq 0.9 Gev^{-2}$ is the universal
     constant, describe the spectrum of  real particles well
     but only approximately. Indeed, the linearity means that the width
     of any resonance is equal to zero;
     the universality of the slope $\alpha^\prime$ is
     connected with the absence of  exotic particles
     \cite{DeAlf}.
     In the meantime we have the stable experimental data on hadronic
     exotics now
     \cite{Landsb}.
     Some of them give  direct information about Regge trajectories
     with slopes $\alpha_g\not= 0.9 Gev^{-2}$
     \cite{Igi}. As regards  the form of the trajectory, the linear
     dependence gives a good approximation for light-flavoured mesons
     and baryons only (see, for example
     \cite{Serg}).
     Moreover, the width of any real resonance does not equal
     zero, of course.

     As it seems, the construction of  D4 free string model,
     while taking into account  the small non-linearity of the trajectories
     and the existence of the different slopes, can be very interesting.

    \subsection*{3. The world-sheet geometry.}

     We will define next the adjunct phase  space  ${\cal H}^{ad},$
     the subset ${\bf W}$ and construct the corresponding
     diffeomorfism ${\bf V}\approx{\bf W}$ in accordance with diagram (9).
     The coordinates in the space ${\cal H}^{ad}$, which will be
     introduced in the following section, naturally
     fall into two groups: the finite number of ''external`` variables
     which are transformed as the tensors under the Lorentz
     transformations of the space-time $E_{1,3}$ and some
     ''internal`` scalar variables.
     In order to define these quantities,
      we  consider in this section the  geometrical construction
     which is quite natural for the studied model. Some parts of the
     section will be analogous to corresponding parts in the
     works
     \cite {Tal1,Tal2}, so that we  give some formulae
     without detail proof.

     Let the constant $\kappa$ is the constant existing
     for given fields $X_\mu$ and $\Psi_\pm$ in accordance with
     the conditions (5). We introduce the tensors

     $$L_\pm^\mu =
{1\over\kappa}\overline\Psi_\pm\Gamma^\mu\Psi_\pm,\qquad
     G^{\mu\nu}_\pm =
{i\over{2\kappa}}\overline\Psi_\pm\left(\Gamma^\mu\Gamma^\nu -
     \Gamma^\nu\Gamma^\mu\right)\Psi_\pm,$$
     which carry the full information about Majorana spinors
     $\Psi_\pm$ and satisfy the properties
     $L_\mu G^{\mu\nu}=0,$~$\quad  L^\mu  L^\nu   =   G^\mu_\rho
     G^{\rho\nu}.$
     After  that we define the pair of vectors  $N_\pm^\mu$:
     $$N_\pm^\mu = {1\over\kappa}\partial_\pm X^\mu +
     {{i\Theta}\over{2\kappa^2}}\overline\Psi_\pm\partial_\pm\Psi_\pm
L_\pm^\mu.$$
     According to the equalities (2) and (5) these
     vectors are light-like and satisfy the conditions
     \begin{equation}
     L_\pm^\mu N_{\mu\pm}=\pm{1\over 2}.
     \end{equation}
     Let us define eight vectors
      ${\bf e}_{\mu\pm},$~~
      $(\mu=0,\dots,3)$:
     $$\left({\bf
e}_{0\pm}\right)^{\mu}=L^{\mu}_{\pm}\pm N_{\pm}^{\mu};\qquad
\left({\bf e}_{3\pm}\right)^{\mu}=\pm
L_{\pm}^{\mu}-N_{\pm}^{\mu};$$
$$\left({\bf e}_{1\pm}\right)^{\mu}=\mp
2G_{\pm}^{\mu\nu}N_{\pm\nu}; \qquad       \left({\bf
e}_2\right)^{\mu}=\varepsilon^{\mu\nu\lambda\rho}\left({\bf
e}_3\right)_{\nu}\left({\bf
e}_0\right)_{\lambda}\left({\bf e}_1\right)_{\rho}.$$
     The direct verification allows to state that these vectors
     give the pair of the orthonormal
     bases in space $E_{1,3}$
     for every point $(\xi^0,\xi^1)$ of  the world-sheet.
     Further it is more convenient to deal with
     the vector-matrices
$${\bf E}_{\pm}= {\bf e}_{\pm}^\mu\sigma_\mu \qquad(\sigma_0=1)$$
     instead the bases  ${\bf e}_{\mu\pm}.$ So, we  define
     $SL(2,C)$-valued chiral field $K=K(\xi^0,\xi^1)$ by means of
     the formula
     \begin{equation}
  {\bf E}_- = K{\bf E}_+K^+.
\end{equation}
   The free equations of motion for original fields $X_\mu$
     and $\Psi_\pm$ lead to the ''conservation laws``
     $\partial_\pm{\bf E}_{\mp}=0$.
    As the consequence, we have the equation for
    chiral field $K=K(\xi^0,\xi^1):$
    \begin{equation}
\partial_-\left(K^{-1}\partial_+K\right)=0.
\end{equation}
     It is special case for well-known Wess-Zumino-Novikov-Witten
     equation
     \cite {Nov,Wit}.
     The left and right currents
     $$Q_-=-(\partial_-K)K^{-1},\qquad Q_+=K^{-1}(\partial_+K)$$
     for the defined chiral field $K$ satisfy the equations
      $\partial_\mp Q_\pm = 0.$
     As can be proven with help of the boundary conditions
     for original string variables ${\bf X}$ and $\Psi,$
     the $sl(2,C)$-valued function
$$Q(\xi)=\cases{Q_+(\xi),&  $\xi\in[0,\pi]$,\cr
-\sigma_1Q_-(-\xi)\sigma_1,&  $\xi\in[-\pi,0]$,\cr}$$
is continious and can be extended $2\pi$-periodically
     and continiously throughout the real axis.

    Let us consider the following auxiliary linear system
     with $2\pi$-periodical coefficients:
     \begin{equation}
T^{\prime}(\xi)+Q(\xi)T(\xi)=0.
\end{equation}
This system plays central role for our subsequent considerations.
       This role is conditioned by the possibility of
     reconstruction of original string variables
     $\partial_\pm X_\mu$ and $\Psi_\pm$ through the matrix-solution
     $T(\xi)$ of the system (13).

     Indeed, the chiral field $K(\xi^0,\xi^1)$ can be written in
     the form
     \cite{Wit}:
      \begin{equation}
     K(\xi^0,\xi^1)=T_-(\xi_-)T_+^{-1}(\xi_+),
     \end{equation}
      for some matrices  $T_\pm\in {\rm SL}(2,C)$.
      Using the definition of the field $K$, we have the
     formulae
\begin{equation}
{\bf E}_\pm(\xi_\pm)=
{\rm T}_\pm(\xi_\pm){\bf E}_0{\rm T}^{\dagger}_\pm(\xi_\pm),
\end{equation}
        where ${\bf E}_0={\bf t}^\mu\sigma_\mu$ is the
     matrix representation of the stationary basis ${\bf t}^\mu.$
     In accordance with the definition of matrix $Q$, we have the
     equalities
     $$T_+(\xi_+)=T(\xi_+);\qquad
     T_-(\xi_-)=i\sigma_1T(-\xi_-),$$
     where $T(\xi)$ is matrix-solution of the system (13) such that
     condition $\det T=1$ holds.
     So, we can reconstruct the vector-matrices ${\bf E}_\pm(\xi_\pm)$
     through the matrix $T(\xi).$ If the constant $\kappa$ is given,
     we can reconstruct the original string variables too.
     Thus the following one-to-one correspondence takes place:
     \begin{equation}
     {\bf V}/E_{1,3}\approx \left(T(\xi),\kappa\right),
     \end{equation}
     where as $E_{1,3}$ we denote the group of the translations
     ${\bf X}\rightarrow{\bf X+A}.$
     The evident formulae for the reconstruction of original string
     variables can be deduced analogously just as in the  works
     \cite {Tal2}
     \footnote{It is important that resulting dependence from  the
      ''variable``   $\kappa$ differs here from the dependence in
     cited work}. For example, we have for the matrices
    $ \partial_\pm\widehat X(\xi_\pm)\equiv
    \partial_\pm X^\mu\sigma_\mu$:
     \begin{equation}
     \partial_\pm\widehat X(\xi_\pm)=\pm
    T^+(\pm\xi_\pm)R(\pm\xi_\pm)T(\pm\xi_\pm),
     \end{equation}
     where the matrix
      $R(\xi)={\rm diag}(\kappa, -2\Theta{\rm Re}Q_{21}(\xi)).$
     If we select the Weyl representation for $\Gamma$-matrices,
     the explicit expressions for reconstructed spinors are quite
     simple:
     \begin{equation}
     \Psi_\pm=\sqrt{\kappa}\left(\matrix{\varphi_\pm\cr
     -\sigma_2\varphi_\pm^\ast\cr}\right),
     \end{equation}
     where
     $\varphi_\pm(\xi_\pm)=\left(\matrix{t_{21}(\pm\xi_\pm)\cr
     t_{22}(\pm\xi_\pm)\cr}\right)$ are the Weyl spinors which were
     expressed in
     terms of the elements $t_{ij}$ of the matrix-solution $T(\xi).$

     This section is finished by the following important statement
     \cite {Tal1}.
     The boundary conditions for original variables ${\bf X}$
     and $\Psi$  were fulfilled, if the equality
\begin{equation}
{\cal M}_1(Q)= \epsilon 1
\end{equation}
     holds for monodromy matrix ${\cal M}_1$ of the
     system (13)  defined in accordance with the equality
     $T(\xi+2\pi)=T(\xi){\cal M}_1.$

     Thus we have as the result of this section the fact that
      the  variables $\partial_\pm X_\mu$ and
     $\Psi_\pm$, constrained by the
     conditions (2) and (5),
     can be reconstructed through the matrix $T$ -- the matrix-soluton
     of the  linear $2\pi$-periodical system
     (13); moreover it is need that the coefficients $Q_{ij}$
     of this system were constrained by equality (19).

    \subsection*{4. The topological charge \\
    and the definition  of space ${\cal H}_1$.}
     In this section we  determine the  adjunct phase space for
     the dynamical system (1).
     The starting point
     of our subsequent consideration is the correspondence (16).
     Note that the matrix-solution $T(\xi)$ is defined up to
     within the transformations
     $$T(\xi)\longrightarrow \tilde T(\xi)=T(\xi)B,$$
     where constant matrix $B\in SL(2,C).$ It is clear
     from the formulae (17) and (18) that these
     transformations are the Lorentz transformations of space-time
     $E_{1,3}.$ Thus we can write for every solution of the system (13):
     \begin{equation}
      T(\xi)=T_0(\xi)B_1(q_1,\dots,q_6),
     \end{equation}
     where the values $q_i$  parametrize the group $SL(2,C)$ somehow
     or other  and the matrix $T_0$  is defined
     from the functions $Q_{ij}(\xi)$ by some unique manner.
     In order to give the correspondent definition of the
     matrix $T_0$, let us fulfill
     the Iwasawa expansion for the matrix-solution $T(\xi):$
     $$T={\cal NEU},$$
      where  ${\cal U}\in{\rm SU}(2)$
     and the matrices  ${\cal E}$ and ${\cal N}$ are following
     $${\cal E}={\rm diag}\left(e^{d}, e^{-d}\right),\qquad
     {\cal N}=\left(\matrix{1&f\cr 0&1\cr}\right).$$
     After that we define the functions $j_a=j_a(\xi),\quad a=1,2,3$:
     $$j_a=-i{\rm Tr}\sigma_a
     [{\cal G}^{-1}Q{\cal   G} + {\cal  G}^{-1}{\cal G}^{\prime}],$$
     where ${\cal G=NE}.$
     Then the matrix ${\cal U}$ satisfies the following linear system:
     \begin{equation}
     {\cal U}^\prime+
     {i\over 2}\left(\sum_{a=1}^{3}\sigma_aj_a\right){\cal U}=0.
     \end{equation}
     Because ${\cal U}\in{\rm SU}(2),$ the functions $j_a(\xi)$
     are the real functions.
       It is more convienent to replace the function $d(\xi),$
      which defines the matrix ${\cal E},$ with the function
      $j_0(\xi)\equiv d^\prime(\xi)$
     and the constant $d_0=d(0).$

     We postulate the following six conditions to fix the matrix $T_0$:
     $$\int_0^{2\pi}f(\xi)d\xi=0,\qquad  d_0=0,\qquad
     {\cal U}(0)={\bf 1}.$$
     Thus we define four real $(j_a)$ and one complex $(f)$ function
     such that the correspondence $Q\leftrightarrow(j_a;f)$ is
     one-to-one. Let us rewrite the condition (19) in terms of
     the introduced functions. So, the matrix $T_0(\xi)$ will be
     $2\pi$-periodical if the functions $f(\xi)$, $j_0(\xi)$ are
     periodical and the equalities
     $$\int_0^{2\pi}j_0(\xi)d\xi=0,\qquad
     {\cal U}(\xi+2\pi)=\epsilon{\cal U}(\xi)$$
     hold. The last equality means that the monodromy matrix
     ${\cal M}$ for linear system (21) satisfies the condition
     $${\cal M}=\epsilon{\bf 1}.$$
     This constraint on the variables $j_a$ leads to the appearance
     of the topological charge $n$ in our model. Indeed, let us consider
     the spectral task
     $${\cal U}^\prime+
     {{i\lambda}\over 2}\left(\sum_{a=1}^{3}\sigma_aj_a\right){\cal U}=0.$$
     The condition (21) holds if and only if this task
     has a  point $\lambda_n=\lambda_n[j_a]$
     of the periodical or antiperiodical spectrum such that
     $\lambda_n=1$  for certain number $n.$
     The equivalent form of this condition is following:
     \begin{equation}
     \Phi_1^m\equiv{\rm arccos}
     \left({1\over 2}{\rm Tr}{\cal M}\right)-\pi m=0.
     \end{equation}
      Thus  we state the  one-to-one correspondence
     \begin{equation}
     {\bf V}/E_{1,3}\approx
     \left( f(\xi), j_0(\xi),\dots,j_3(\xi);
     ~q_1,\dots,q_6; ~\kappa\right).
     \end{equation}
     The whole number $n$
     is the topological charge in our theory;
     the continious deformation of the string configuration
     $(f(\xi),\dots)$ for some $n$ into the configuration
     $(f(\xi),\dots)$ with other number $m$
     breaks either boundary conditions or gauge (5).

     Our following step is to define six parameters $q_i$ according
     to the representation (20). Moreover, we must to add four constants
     $Z_\mu$  for the reconstruction of the variables
     $X_\mu$ from the derivatives $\partial_\pm X_\mu.$
     Let us consider the usual Noether expressions for the energy-momentum
     $P_\mu$ and the moment $M_{\mu\nu}:$
     $$P_\mu=  {1\over{4\pi\alpha^\prime}}
     \int_0^{\pi}\dot X_\mu d\xi^1,$$
     $$M_{\mu\nu}= {1\over{4\pi\alpha^\prime}}
     \int_0^\pi\left(X_\mu\dot X_\nu-X_\nu\dot
     X_\mu\right)d\xi^1-{{i\Theta}\over{8\pi\alpha^\prime}}
     \sum_{\epsilon=\pm}\int_0^\pi
     \overline\Psi_\epsilon\left(\Gamma_\mu\Gamma_\nu-
     \Gamma_\nu\Gamma_\mu\right)\Psi_\epsilon d\xi^1.$$
     Let
     $w_{\mu}=-(1/{2})
     \varepsilon_{\mu\nu\lambda\sigma}M^{\nu\lambda}P^{\sigma}.$
     In accordance with the formulae (17) and (18) we have the
     equalities:
     \begin{equation}
     (P)^2=\left({\Theta\over{4\pi\alpha^\prime}}\right)^2
     \sum_{l=0}^2\left(\kappa\over\Theta\right)^l D_l,
     \end{equation}
     \begin{equation}
     (w)^2={\Theta^6\over({4\pi\alpha^\prime})^4}
     \sum_{l=0}^6 \left(\kappa\over\Theta\right)^l F_l.
     \end{equation}
     It is important that the coefficients $D_l$ and $F_l$
     in the polynomials (24) and (25) depend on the functions
     $f(\xi)$ and $j_a(\xi)$ only.
     This fact means that these formulae give the
     $\kappa$-parametric form
     of ''constraint``
     \begin{equation}
     \Phi_2(P^2,w^2;f,j_0,\dots,j_3)=0.
     \end{equation}

     The main idea is to use the components $P_\mu$ and $M_{\mu\nu}$
     as an additional variables instead the constants $Z_\mu,$~
     $q_i$ and $\kappa.$
     The exact statement  is following.
     \begin{pp}

     { Let two-parametric group $G_2$ was composed from the
     transformations:
     \begin{enumerate}
     \item{ rotations
     $X_\mu\rightarrow \Lambda_\mu^\nu(\phi)X_\nu$ in the
     space-like plane which is orthogonal with the vector
     $P_\mu$ and pseudo-vector $w_\mu;$}
      \item{translations $X_\mu\rightarrow X_\mu+cP_\mu.$}
     \end{enumerate}
     Then, if the quantities $f(\xi),$~ $j_a(\xi)$ ($a=0,\dots,3$),
     $P_\mu$ and $M_{\mu\nu}$ are constrained by the  equalities  (22)
     and (26),
     the diffeomorfism
     $${\bf V}/G_2\approx
     (f(\xi), j_0(\xi),\dots,j_3(\xi); P_\mu, M_{\mu\nu})$$
     takes place.}
     \end{pp}
     The sketch of the proof is following. Let the auxiliary vector
     field ${\bf X}_{(0)}$ and the spinor fields $\Psi_{(0)\pm}$
     were defined from the variables $f(\xi)$ and $j_a(\xi)$ with
     help of the formulae (17) and (18), where the replacement
     $T(\xi)\rightarrow T_0(\xi)$ has been fulfilled.
     We define next the vector $P_{(0)\mu}$ and pseudo-vector
     $w_{(0)\mu}$ by means of the replacement
     ${\bf X}\rightarrow{\bf X}_{(0)}$ and
     $\Psi_\pm\rightarrow\Psi_{(0)\pm}$ in the correspondent
     Noether expressions. Let $P_\mu$ arbitrary time-like vector
     and $M_{\mu\nu}$ arbitrary antisymmetrical tensor. Then,
     if the constraint (26) takes place, the matrix $B\in SL(2,C)$
     exists such that the equalities
     $$\hat P=B^+\hat P_{(0)}B,\qquad \hat w=B^+\hat w_{(0)}B$$
     hold. That is why we can reconstruct the matrix $T=T_0B.$
     Moreover, we can restore  the integration constants
     $Z_\mu$ because the full moment $M_{\mu\nu}$ consists the
     information about center of mass of the string.
     Consequently, the original string variables
     ${\bf X}$ and $\Psi_\pm$ can be restored from the
     variables $f,j_a,$
     $P_\mu$ and $M_{\mu\nu}.$
     More detail investigations show that this reconstruction
     will be smooth and two-parametric
     arbitrariness exists, so that the corresponding cosets
     appear.

     To describe the degrees of the freedom connected with the
     group $G_2,$ we introduce the additional coordinates
     $q$ and $\theta,$ such that $-\infty<q<\infty$ and
     $\theta\in[0,2\pi].$
     Now we give the straightforward definition of the
     adjunct phase space
     ${\cal H}^{ad}$ for the considered string model.
     This is manifold such that any point $M\in{\cal H}^{ad}$ has
     the following coordinates: 1) $2\pi$-periodical complex function
     $f(\xi)$ without zero mode; 2) $2\pi$-periodical real functions
     $j_a(\xi)$ (a=0,1,2,3) such that the   function $j_0$ has not
     zero mode;
     3) 4-vector $P_\mu$ such that the inequality $P^2>0$ holds;
     4) antisymmetrical tenzor $M_{\mu\nu};$
     5) four additional coordinates $q$, $\theta$, $p$ and $\chi.$
     Let us define the  Poisson brackets
     $$\left\{f(\xi),\overline f(\eta)\right\}^0=
     {\alpha^\prime\over{\Theta^2}}\delta^\prime(\xi-\eta),\qquad
     \left\{j_0(\xi),j_0(\eta)\right\}^0=
     -2{\alpha^\prime\over{\Theta^2}}\delta^\prime(\xi-\eta),$$
     $$\left\{j_a(\xi),j_b(\eta)\right\}^0=
     2{\alpha^\prime\over{\Theta^2}}\left(
      -\delta_{ab}\delta^{\prime}(\xi-\eta)+
\varepsilon_{abc}j_c(\xi)\delta(\xi-\eta)\right)$$
(where $a,b,c=1,2,3$ and also
$\delta(\xi)=\sum_n{\rm e}^{in\xi}$),
  $$\left\{M_{\alpha\beta},M_{\gamma\delta}\right\}^0=
g_{\alpha\delta}M_{\beta\gamma}+g_{\beta\gamma}M_{\alpha
\delta}-g_{\alpha\gamma}M_{\beta\delta}-
g_{\beta\delta}M_{\alpha\gamma},$$
   $$\left\{M_{\alpha\beta},P_\gamma\right\}^0=
g_{\beta\gamma}P_{\alpha}-g_{\alpha\gamma}P_{\beta},$$
     $$\{p,q\}^0=1,\qquad \{\chi,\theta\}^0=1$$
     (The other possible brackets are equal to zero).
     With respect to defined brackets
     the space ${\cal H}^{ad}$ is the Poisson manifold.

     The  manifold ${\bf W}$ is defined as follows.
     As first we require, that the equalities
     $$\Phi_3\equiv p=0,\qquad \Phi_4\equiv\chi=0$$
     hold. As $\Phi_1^n$ we denote the
     constraint (22) for some topological number $n.$
     Let the  set ${\bf W}_n\subset{\cal H}^{ad}$ be the
     surface of the constraints $\Phi_i,$~ $i=1,\dots,4,$
     where $\Phi_2=\Phi_2^n.$ Then,
     $${\bf W}=\mathop\cup_{n\in Z}{\bf W}_n.$$

     \begin{pp}
     The constraints $\Phi_i=0$, $i=1,\dots,4$
     will be  first type constraints
     with respect to the brackets $\{\cdot,\cdot\}^0.$
     \end{pp}

     Indeed, $\{\Phi_i,\Phi_j\}=0$ for $i=1,2$ and $j=3,4,$ or
     $i=3$ and $j=4.$ Let us prove that
     $\{\Phi_1,\Phi_2\}\propto\Phi_1.$
     We first note that the matrix ${\cal M}$
     depends on the variables $j_a$, $a=1,2,3$ only.
     Let us calculate the brackets of the matrix elements
     of the matrices ${\cal   M}$  and
     $Q_g=(i/2)\sum_aj_a\sigma_a$.
     The identity
     $$\left\{\bigl({\cal U}^\prime(\xi)+Q_g(\xi){\cal U}(\xi)
     \bigr){\mathop\otimes_,}{\cal M}\right\}^0\equiv 0$$
      holds on space ${\cal H}^{ad},$ therefore we can
     apply for such calculations the Leibniz rule
      $\{AB,C\}^0=A\{B,C\}^0+\{A,C\}^0B$ and the definition
     of the matrix ${\cal M}.$
     As result we have the equality
    $$\left\{Q_g(\xi){\mathop\otimes_,} {\cal M}\right\}^0
=\left[1\otimes{\cal M},C(\xi)\right],$$
   where the square brackets denote the commutator
     $4\times4$ matrices.
     The explicit form of the matrix $C(\xi)$ does not important
     here because it is clear if  ${\cal M}\propto 1,$ than
    $\left\{Q_g(\xi){\mathop\otimes_,} {\cal M}\right\}^0\equiv0.$
    Consequently, we have
     $$\left\{\Phi_1, A\right\}^0\propto\Phi_1$$
    for arbitrary function $A=A(f,j_a;P_\mu,M_{\mu\nu};q,\theta),$
     so that the Corollary is proven.

      The dynamical equations
     $$\left\{H_0,X_\mu\right\}^0=
     {{\partial X_\mu}\over{\partial\xi^0}},\qquad
     \left\{H_0,\Psi_\pm\right\}^0=
     {{\partial\Psi_\pm}\over{\partial\xi^0}}$$
     hold for the hamiltonian
     $$H_0={\Theta^2\over{2\pi\alpha^\prime}}
     \left(\int_0^{2\pi}\vert f(\xi)\vert^2
     d\xi+{1\over 4}\sum_{a=0}^3
     \int_0^{2\pi}j_a^2(\xi)d\xi\right).$$
     These formulae can be proven with help of the representation
     (17) and (18) for original string variables $X_\mu,$~ $\Psi_\pm$
     and with help of the obvious equalities
     $$\left\{H_0,j_a\right\}^0 = j_a^\prime,\qquad
     \left\{H_0,f\right\}^0 = f^\prime,\qquad
     \left\{H_0,P_\mu\right\}^0 =\left\{H_0,M_{\mu\nu}\right\}^0 =0.$$
      It can be verified  directly that all constraints
     in our thery  are co-ordinated with dynamics.

     {\it Remark.} It is clear that the brackets of the variables
     $P_\mu$ and $M_{\mu\nu}$ are motivated by Poincar\'e algebra.
     Consequently, we  have  two  annulators here:
     $P_\mu P^\mu$ and $w_\nu w^\nu.$
     But every Poisson structure
     $\{\cdot,\cdot\}$ must be co-ordinated with the tensor property
     of all considered functions. So, for instance, the equality
     $\{P_\mu,A_\nu\} =g_{\mu\nu}$  must holds for any 4-vector
     $A_\mu$ in order to the dynamical variables $P_\mu$ generate
     Poincar\'e translations.
     In our theory the integration of the formula (17) gives
     the expression for radius-vector $X_\mu(\xi^0,\xi^1):$
     $$
     \widehat X(\xi^0,\xi^1)=\widehat Z+{{\xi^0+q}\over\pi}
     \widehat P          -{i\over\pi}\sum_{n\not=0}{{\widehat
     C_n}\over n}{\rm e}^{in\xi^0}\cos n\xi^1,$$

     $${\rm where}\qquad \widehat C_n=
     \int_0^{2\pi}T^\dagger(x)R(x)T(x){\rm e}^{-inx}dx$$
     and $Z_\mu=M_{\mu\nu}P^\nu/P^2.$
     Therefore, we have the brackets
      $$\{P_\mu,X_\nu\}^0 = g_{\mu\nu}-{{P_\mu P_\nu}\over{P^2}},$$
     which are co-ordinated with the fact that function $P^2$
     will be annulator.
     This  means that  for every constant 4-vector $b^\mu$
     the following formula takes place:
     $${\rm e}^{b^\mu\{P_\mu,\dots}X_\nu=X_\nu+b_\nu-
     \left({{b_\rho P^\rho}\over{P^2}}\right) P_\nu.$$
     Thus, with respect to the defined brackets,
     the variables $P_\mu$  will generate
     the Poincar\'e translations
     on the correspondent cosets   only. Same situation holds
     for the rotations, mentioned in the definition of the group $G_2.$
     The additional constraints $\Phi_3$ and $\Phi_4$ allow to
     reconstruct the correct co-ordination of the introduced Poisson
     brackets with translations and rotations.
     Indeed, let us consider  Lie operator
     $$L^\mu(P)={\bigl\{P^\mu,\dots} +
     \pi{{P^\mu}\over{P^2}}\bigl\{\Phi_3,\dots $$
     instead of the conventional operator of translation
     $\{P^\mu,\dots.$
     In accordance with the definitions of the variable $q$
     and the constraint $\Phi_3$,  the equality
     $${\rm e}^{a_\mu L^\mu(P)} X_\nu=X_\nu+a_\nu$$
     holds.
     Analogously, Lie operator $\{M^{\mu\nu},\dots$ must be
     improved by means of adding the term with the operator
     $\{\Phi_4,\dots.$

                    $$***$$
     Let us discuss the quantization of the suggested model.
     We surmise that the structure of the fundamental Poisson
     brackets algebra ${\cal A}_{cl}$ gives some information about the
     constructed space of the quantum states. In our model
     this algebra has the form
     $${\cal A}_{cl}={\cal A}_{int}\oplus{\cal P},$$
     where ${\cal A}_{int}$ the Poisson brackets algebra of the
     ''internal`` variables $f(\xi)$,  $j_a(\xi)$ and
     ${\cal P}$ is the Poincar\'e algebra. It should be emphasized that
     the energy-momentum and moment of the string (1) are independent
     fundamental variables, so there are no problems with the quantum
     ordering when we construct the quantum generators of Poincar\'e
     transformations.

      The defined new variables
     are complicated functions from the original fields $X$ and
     $\Psi,$
     that is why the correct introduction of quantum fermionic
     fields is not so obvious here.  The following proposition clarifies
     this question
     \cite {Tal2}
     \begin{pp}
      { The equalities $\Psi^A_\pm(\xi)\equiv{\rm const}$ hold
     if and only if the equalities $j_a(\xi)\equiv 0$ for
      $a=0,\dots,3$ take place.}
     \end{pp}
     This statement means that, in spite of the complicated dependence
     of the variables $f$ and $j_a$ from the original variables
     $X_\mu$ and $\Psi,$ the bosonic and fermionic degrees of the
     freedom are still non-mixed. It is natural to fulfill the
     quantization of the variables $j_a$ in terms of the fermionic
     fields with help of the bosonization procedure
     \cite {Wit}. Thus the natural Hylbert space of the
     of the quantum states of the string wil be following:
     $${\bf H}=\mathop\oplus_{l,i,s}\Bigl({\bf H}_b\otimes
     {\bf H}_f\otimes{\bf H}_{\mu^2_i,s}\Bigr),$$
     where the spaces ${\bf H}_{\mu^2,s}$ are the spaces
     of irreducible representations of Poincar\'e algebra
     ${\cal P},$ labeled by the eigenvalues of the Cazimir
     operators  $P^\mu P_\mu$ and $w^\mu w_\nu;$
     ${\bf H}_b-$ the Fock space of two-dimensional  bosonic
     field in the ''box`` and ${\bf H}_f$ -- the Fock space of
     two-dimensional fermionic field in the ''box``.
     The corresponding physical vectors of states must be selected
     with help of the ''Shr\"odinger equations''
     $$\Phi_i\mid\psi_{phys}\rangle=0,$$
     where $\Phi_i$ are the quantum expressions for considered
     constraints.

      The other consequense of this Corollary
     is that   the suggested theory can be considered
     as the new spinning generalization of the standard bosonic
     string model with   the light-cone gauge. Indeed,
     the standard  light-cone  gauge for bosonic string can be written
in the form
     \begin{equation}
     n_\mu\partial_\pm X^\mu=\pm p_+/2,
     \end{equation}
where light-like vector $n_\mu$ are selected usually as $(1,0,0,1).$
     In our case both spinors $\Psi_\pm$ are Majorana spinors in
     $D4$ space-time, so the vectors
     $n^\mu_\pm=\overline\Psi_\pm\Gamma^\mu\Psi_\pm$
     will be light-like always.
     The reduction $j_a\equiv 0$ means  that these vectors are
     constant,
     moreover $n_+^\mu=n_-^\mu $ in accordance with the usual boundary
conditions for the spinor variables. Therefore, we have the theory
     with the gaude (27) where the light-like vector $n_\mu$
     constant, but arbitrary. If  we require  $\Theta=0,$ the action
     (1) takes the standard bosonic form.
     The real and imaginary parts of the functions $f(\pm\xi_\pm)$
     will be the (well-known) transversal components for vectors
     $\partial_\pm{\bf X}.$
     With respect to the formulae (5)
     $$\Omega_{ij}d\xi^i d\xi^j\propto \kappa^2
     [(d\xi_+)^2 +(d\xi_-)^2],$$
     so this form has a good limit when $\Theta\rightarrow 0.$
      In spite of this fact we assume  that the two-metrics
     $\Omega_{ij}$ does not natural object for bosonic case $\Theta=0,$
     because the  spinor variables are absent here.  This case was
     studied  resently in the author's work
     \cite{Tal3},
     where both classical and quantum version of the model
     investigated in detail.  As result, we have Regge trajectories
     $\hbar\sqrt{s(s+1)}=\alpha_n \mu^2,$ where the slopes
     $\alpha_n,$ ~~$n=1,2,\dots$ are the eigenvalues for some spectral
     task in the space of quantum states.   The case $j_a\equiv 0$,
     but $\Theta\not= 0$ is quite similar technically, but it is more
     interesting, because leads to the more complicated trajectories.

     Note that we can fulfill  some unusual reduction $f\equiv 0$
     in our model which corresponds to the string, where all bosonic
     degrees of the freedom are ''frozen``. Previous investigation
     of this case was made in the work
     \cite {Tal1}, where the quantization was discussed too. This case
     more complicated because (nontrivial) topological condition (22).
     Author hopes to study the general quantum case in the future.

    \subsection*{5. Concluding remarks.}

     In this paper we suggest new consept of adjunct phase
     space to investigate the open spinning string.
     It should be stressed that suggested approach leads to D4
     covariant theory both in the classical and in the
     quantum cases. Main result is new non-trivial
     Regge spectrum which can be applied in our opinion to
     the description of the exotic particles.   The
     dependence  $J=J(P^2),$ where the spin
     $ J=\sqrt{w^2/P^2},$ can be analysed already on classical
     level with help of the formulae (24) and  (25).
     It will be essentially non-linear for small
     masses although for large $P^2$ we have the asymptotics
     $J\propto P^2+{\cal O}(\sqrt{P^2}).$

     Let us note that we have two fundamental constants in the theory:
     $\alpha^\prime$ and $\Theta.$ Because the spinor part of the
     action (1) vanishes on the equations of the motion, the constant
     $\Theta$ can be introduced in the model not as the fixed constant
     but as the additional variable. The previous investigation
     of the theory with the original
     configuration space $(X,\Psi;\Theta)$ instead of the space
     $(X,\Psi)$ was fulfilled in the works
     \cite{Tal1,Tal2}.
     It should be stressed that such  extension leads to the
     scale-invariant theory if we define the scale transformations
     as $(X,\Psi;\Theta)\rightarrow$~ $(aX,\sqrt{a}\Psi;a\Theta).$
     As a natural result, here the linear dependence  $J\equiv\sqrt{s(s+1)}=
     \alpha_n\mu^2$ was deduced: we have the set of Regge trajectories
     with zero intercepts but with various slopes $\alpha_n.$
     We consider the value $\Theta$  as the constant but not as the
     variable in this work. Because the scale invariance is broken in
     this case the resulting Regge spectrum is more complicated than
     the spectrum in the articles
     \cite{Tal1,Tal2}.
     Note that the  models of bosonic strings with non-standard spectrum
     were suggested last time in the works
     \cite{Barb,SolLD}.

     It is known that the slope $\alpha^\prime$
     of the Regge trajectory can be
     connected with the tension $\tau$ of the string:
     $\alpha^\prime\propto{1/\tau}.$
     If the tension is constant  and there are
     no other internal forces, the slope will be constant too.
     Thus the complicated form of the Regge trajectory bears
     a relation, probably, to some additional internal
     forces within the string.
     Note that the models, where spinning degrees of the freedom
     were connected with distributed charges and currents,
     were investigated early (see, for example,
     \cite {LarNel}).
     It appears that  such interpretation is possible in our case too.
     The interesting moment here is that the model has topological
     charge $n$ which vanishes if the fermionic variables disappear.

     This work was supported by the Russian Foundation for
     Basic Reseach, Project No. 96-01-00299.

   \end{document}